\documentclass[a4paper,11pt]{article}
\pdfoutput=1 
\usepackage{jcappub}
\usepackage{amsmath}
\usepackage{graphicx} 
\usepackage{latexsym}
\usepackage{xspace}
\usepackage{color}
\usepackage{hyperref} 
\usepackage{bm}
\usepackage{relsize}
\usepackage{tabularx}
\usepackage{multirow}
\usepackage{amssymb}
\usepackage[table]{xcolor}
\usepackage{slashbox}
\usepackage{braket}

\usepackage[utf8]{inputenc}
\usepackage[normalem]{ulem}

\usepackage{natbib}

\allowdisplaybreaks

\makeatletter
\gdef\@fpheader{}
\g@addto@macro\bfseries{\boldmath}
\makeatother

\makeatletter
\renewcommand*\env@matrix[1][\arraystretch]{%
  \edef\arraystretch{#1}%
  \hskip -\arraycolsep
  \let\@ifnextchar\new@ifnextchar
  \array{*\c@MaxMatrixCols c}}
\makeatother

\newcommand{\R}{{\cal R}}
\newcommand{\Pnad}{P_{\mathrm{nad}}}
\newcommand{\Pad}{P_{\mathrm{ad}}}
\newcommand{\dPnad}{\delta \Pnad}

\newcommand{\ovl}[1]{\overline{#1}}

\newcommand{\ie}{i.e.\xspace}

\newcommand{\eg}{e.g.\xspace}

\newcommand{\dd}{\mathrm{d}}

\newcommand{\sss}[1]{{\scriptscriptstyle{#1}}}

\newcommand{\uPl}{\mathrm{Pl}}

\newcommand{\usssPl}{\sss{\uPl}}

\newcommand{\cs}{c_{_\mathrm{S}}}

\newcommand{\Mp}{M_\usssPl}

\newcommand{\beq}{\begin{equation}}
\newcommand{\eeq}{\end{equation}}
\newcommand{\bea}{\begin{eqnarray}}
\newcommand{\eea}{\end{eqnarray}}

\newlength{\wsingfig}
\newlength{\wdblefig}
\newlength{\wquadfig}
\newlength{\wtriplefig}
\newcommand{\Eq}[1]{Eq.~(\ref{#1})}
\newcommand{\Eqs}[1]{Eqs.~(\ref{#1})}

\newcommand{\Refc}[1]{Ref.~{\cite{#1}}}

\newcommand{\Sec}[1]{Sec.~\ref{#1}}

\newcommand{\App}[1]{Appendix~\ref{#1}}

\usepackage[modulo]{lineno}

\title{On the Hamilton-Jacobi approach to inflation beyond slow roll}

\author[a]{Danilo Artigas,}
\emailAdd{artigas@tap.scphys.kyoto-u.ac.jp}

\author[b]{Emmanuel Frion,}
\emailAdd{efrion@uwo.ca}

\author[c]{Tays Miranda,}
\emailAdd{taysmiranda@id.uff.br}

\author[d]{Vincent Vennin}
\emailAdd{vincent.vennin@ens.fr}

\author[e]{and David Wands}
\emailAdd{david.wands@port.ac.uk}

\affiliation[a]{Department of Physics, Kyoto University, Kyoto 606-8502, Japan}
\affiliation[b]{Department of Physics and Astronomy, Western University, N6A 3K7, London, Ontario, Canada}
\affiliation[c]{CBPF - Centro Brasileiro de Pesquisas Físicas, 22290-180, Rio de Janeiro, RJ, Brazil}
\affiliation[d]{Laboratoire de Physique de l’Ecole Normale Supérieure, ENS, CNRS, Université PSL, Sorbonne Université, Université Paris Cité, F-75005 Paris, France}
\affiliation[e]{Institute of Cosmology \& Gravitation, University of Portsmouth, Dennis Sciama Building,
Burnaby Road, Portsmouth, PO1 3FX, United Kingdom}

\date{today}

\begin{document}
\begin{flushright}
\small{\textsc{Matches published version in JCAP}}
\end{flushright}

\sloppy

\abstract{The Hamilton-Jacobi approach is a powerful tool to describe super-Hubble dynamics during cosmological inflation in a non-linear way. A key assumption of this framework is to neglect anisotropic perturbations on large scales.
We show that neglecting the anisotropic sector in the momentum constraint corresponds to discarding the non-adiabatic mode of scalar-field perturbations at large scales.
Consequently, the Hamilton-Jacobi approach cannot be used to describe the evolution of large-scale perturbations during inflation beyond slow roll, when non-adiabatic fluctuations play an important role on super-Hubble scales due to the absence of an attractor trajectory.
As an example, we analyse the case of cosmological perturbations during a phase of ultra-slow-roll inflation.}



\maketitle

\flushbottom

\section{Introduction}
\label{sec:intro}

Scalar fields play a central role in current models of the very-early high-energy evolution of our universe. A canonical scalar field, $\phi$, with self-interaction potential, $V(\phi)$, can drive an accelerated expansion when its potential dominates over its kinetic and gradient energy densities~\cite{Ellis:2023wic}. This inflation can lead to a homogeneous and isotropic patch described at least approximately by a spatially-flat Friedmann-Lema\^itre-Robertson-Walker (FLRW) metric~\cite{Aurrekoetxea:2024mdy}. 

Quantum fluctuations inevitably lead to local variations in the scalar field and a period of slow-roll inflation naturally gives rise to an almost scale-invariant distribution of primordial density perturbations after inflation.
However, there has recently been renewed interest in inflation models where a brief violation of slow-roll enables a strong enhancement of the comoving curvature perturbation on small scales~\cite{Garcia-Bellido:2017mdw,Motohashi:2017kbs,Germani:2017bcs,Ballesteros:2017fsr}. 
These amplified perturbations can lead to interesting consequences such as an induced gravitational-wave background~\cite{Acquaviva:2002ud,Ananda:2006af,Baumann:2007zm,Domenech:2021ztg} or
the formation of primordial black holes, which arise from the collapse of sufficiently large over-densities after inflation \cite{Zeldovich:1967lct,Hawking:1971ei,Carr:1974nx,Escriva:2022duf,LISACosmologyWorkingGroup:2023njw}.

A homogeneous scalar field, $\phi(t)$, in an FLRW spacetime obeys the Klein-Gordon equation
\bea
\label{eq:KG}
\ddot{\phi}+3H\dot{\phi}+V_{,\phi}=0\, ,
\eea
where $V_{,\phi}$ is the derivative of the potential with respect to $\phi$ and a dot denotes differentiation with respect to cosmic time. The Friedmann constraint equation
\bea 
\label{eq:Friedmann}
H^2=\frac{\frac{\dot\phi^2}{2}+V(\phi)}{3\Mp^2}
\eea 
gives the Hubble expansion rate, $H:=\dot{a}/a$, where  $a$ is the scale factor. $\Mp$ is the reduced Planck mass and we work with natural units such that $\hbar=c=1$.
The Klein-Gordon equation \eqref{eq:KG} is a second-order differential equation, leading to two arbitrary integration constants in the general solution. Thus the density $\rho=\dot\phi^2/2+V(\phi)$ and pressure $P =\dot\phi^2/2-V(\phi)$ can vary independently. In particular, small fluctuations about a reference background solution admit both adiabatic pressure perturbations, $\delta \Pad= \cs^2\delta\rho$, and non-adiabatic perturbations,  
\bea
\delta \Pnad:= \delta P - \frac{\dot{P}}{\dot{\rho}} \delta\rho \,,
\label{def:Pnad}
\eea
where the adiabatic sound speed is given by the background trajectory, $\cs^2:=\dot{P}/\dot\rho$. Throughout this paper we will identify non-adiabatic perturbations in scalars $X$ and $Y$ with
\bea
\label{def:Xnad}
\delta X_{Y} := \delta X - \frac{\dot{X}}{\dot{Y}} \delta Y \,,
\eea
generalising \Eq{def:Pnad}. Thus adiabatic fluctuations correspond to perturbations aligned with the background trajectory in the background phase space~\cite{Wands:2000dp,Gordon:2000hv}, \ie, corresponding to an adiabatic evolution.

For adiabatic perturbations in a scalar-field-dominated cosmology the comoving curvature perturbation, $\R$, is conserved at first order in the fluctuations~\cite{Kodama:1984ziu,Wands:2000dp,Weinberg:2003sw,Malik:2008im,Romano:2015vxz} and it can be shown that~\cite{Jackson:2023obv}
\begin{equation}
\label{eq:dotR}
    \dot{\R} = \frac{3H^2}{2\dot{V}} \delta \Pnad \,.
\end{equation}
At the same time, each wavemode, $\R_k$, with comoving wavenumber $k$, obeys a second-order equation~\cite{Kodama:1984ziu}
\begin{equation}
\label{eq:Rkeom}
    \R_k''+2\frac{z'}{z} \R_k' + k^2 \R_k = 0 \,,
\end{equation}
where $z:= a\dot\phi/H$ and a prime denotes a derivative with respect to conformal time, $\dd / \dd \eta := a(\dd / \dd t)$.
In the long-wavelength limit, $k\to0$, \Eq{eq:Rkeom} has the general solution
\begin{equation}
\label{eq:R0}
    \R^{(0)}(\eta) = C + D \int_{\eta_*}^\eta \frac{\dd \tilde\eta}{z^2(\tilde\eta)} \,.
\end{equation}
Comparing \Eq{eq:R0} with \Eq{eq:dotR}, we identify $\R^{(0)} = C$ with the leading-order term in a gradient expansion of the adiabatic mode.

The time-dependent term in \Eq{eq:R0}, proportional to $D$, describes a non-adiabatic perturbation.
It is the decaying mode if we set $\eta_*=\eta_{\rm end}$ at the end of inflation.
In single-field inflation this term is often neglected on large scales (typically taken to be wavelengths larger than the Hubble scale, \ie modes with comoving wavenumber $k\ll aH$)~\cite{Salopek:1990jq,Salopek:1990re,Prokopec:2019srf,Rigopoulos:2021nhv}. This is expected to be a good approximation in situations where there is an attractor trajectory in the homogeneous phase space for the scalar field. 
This is because, in such cases, large-scale fluctuations join the attractor and thus become aligned with the background solution, \ie they are adiabatic.
For example, in slow-roll inflation, strong Hubble damping leads to a quasi-equilibrium solution for the scalar field where the damping term in the Klein-Gordon equation~\eqref{eq:KG}, $3H\dot\phi$, is equal and opposite to the potential gradient, $V_{,\phi}$, as the field rolls down its potential. This slow-roll solution is an attractor for a sufficiently flat potential, allowing one to neglect the non-adiabatic perturbation on super-Hubble scales. 

While it has been known for some time that violation of slow roll can lead to rapid growth of the curvature perturbation on super-Hubble scales~\cite{Leach:2001zf,Kinney:2005vj}, there is some confusion as to whether the existence of non-adiabatic perturbations in single-field inflation is consistent with the long-wavelength limit of cosmological-perturbation theory, and in particular with the use of the separate-universe approach, in which long-wavelength perturbations are described as local perturbations of the homogeneous and isotropic equations of motion~\cite{Lifshitz:1960, Starobinsky:1982ee, Salopek:1990jq, Comer:1994np, Sasaki:1995aw, Sasaki:1998ug, Wands:2000dp, Lyth:2003im, Rigopoulos:2003ak, Lyth:2004gb, Lyth:2005fi,Cai:2018dkf,Cai:2022erk,Pi:2022ysn}. For a dissenting viewpoint, see~\cite{Cruces:2018cvq,Cruces:2021iwq,Cruces:2022dom}.

One approach to determining the large-scale dynamics during inflation is the Hamilton-Jacobi approach \cite{Salopek:1990jq,Rigopoulos:2003ak}. This consists of solving the Hamilton-Jacobi equation, which is obtained by injecting the solution to the momentum constraint into the Hamiltonian constraint of general relativity. However, as noted in~\cite{Tomberg:2022mkt}, at the moment of solving the constraints the anisotropic part of the extrinsic curvature has already been neglected assuming it decays quickly on large scales~\cite{Salopek:1990jq,Rigopoulos:2003ak}. As a consequence, what is referred to as the ``momentum constraint'' is actually just its isotropic part. The main result 
of this paper is to show that, by setting the isotropic part of the momentum constraint to zero in single-field inflation, one automatically discards the non-adiabatic pressure perturbation and, therefore, may fail to capture the large-scale dynamics beyond slow roll. 

In \Sec{sec:HJ}, we summarise the main ideas behind the Hamilton-Jacobi approach, and explain why perturbations of the local expansion rate are proportional to perturbations of the scalar field in this setup. In \Sec{sec:pert}, we briefly recap linear-perturbation theory and introduce the gradient expansion. In \Sec{sec:dpnad}, we introduce the non-adiabatic pressure perturbation and show how it is related to the isotropic part of the momentum constraint. In \Sec{sec:appli}, we explicitly solve the dynamics in the context of linear perturbations to obtain $\delta\Pnad$ both in slow roll and ultra-slow roll (USR). We show that, in the spatially-flat gauge, $\delta\Pnad$ does not vanish in USR at leading order in a gradient expansion. As such, the Hamilton-Jacobi approach may not be applicable in this context. We present our conclusions in \Sec{sec:conc}.

\section{Hamilton-Jacobi approach}
\label{sec:HJ}

Let us start by reviewing the Hamilton-Jacobi approach. A detailed derivation of the equations given below will follow, our goal in this section is to summarise the problem in simple terms.
General relativity is subject to diffeomorphism invariance, hence it comes with constraints and Lagrange multipliers, corresponding to the freedom in the choice of the coordinate system. When cosmological perturbations are expanded around a homogeneous and isotropic background, described by the FLRW metric with a homogeneous scalar field as the matter content, these reduce to two constraints in the scalar sector, the energy constraint and the momentum constraint. Metric perturbations can be divided into isotropic perturbations and anisotropic ones~\cite{Langlois:1994ec}. 
When only isotropic degrees of freedom are considered, the momentum constraint reduces to~\cite{Salopek:1990jq}
\bea
\label{eq:MomentumConstraint:SalopekBond}
\partial_i  H = -\frac{1}{2\Mp^2}\dot{\phi}\partial_i\phi\, ,
\eea 
where $H$ is the local expansion rate. This equation will be derived in detail in \Sec{sec:pert}, while here we focus on its implications. 

Dividing the field, $\phi$, and expansion rate, $H$, into a background, homogeneous component, and a perturbation component, $H(\vec{x},t)= {H}(t)+\delta H(\vec{x},t)$ and $\phi(\vec{x},t)= {\phi}(t)+\delta \phi(\vec{x},t)$, it is clear that the momentum constraint \eqref{eq:MomentumConstraint:SalopekBond} is a statement about inhomogeneous perturbations only,
\bea
\label{eq:deltaH:deltaphi}
  \delta H = -\frac{1}{2\Mp^2}\dot{\phi}\delta \phi\, .
\eea 
This implies that $\delta H$, which a priori depends on all perturbed variables, is in fact a function of $\delta \phi$ only,
\bea
\label{eq:f:def}
\delta H(\vec{x},t) = f[\delta\phi(\vec{x},t),t]\, .
\eea 
Moreover, at the background level, combining the Friedmann and Klein-Gordon equations, one can readily show that $\dot{{H}}(t)=-\dot{{\phi}}^2(t)/(2\Mp^2)$. Together with \Eq{eq:deltaH:deltaphi}, this implies that $\dot{{H}}(t) \delta\phi  = \dot{{\phi}}(t) \delta H$, hence the vector $(\delta\phi,\delta H)$ is aligned with $(\dot{{\phi}}(t),\dot{{H}}(t))$ in phase space and these perturbations are adiabatic in the sense given in \Eq{def:Xnad}.
This is why \Eq{eq:f:def} can be promoted into a relationship between the full fields,
\bea
\label{eq:f:adiab}
 H(\vec{x},t) = g[\phi(\vec{x},t)]\, .
\eea
This relation is the starting point of the Hamilton-Jacobi formalism, and it has been used to track the evolution of adiabatic perturbations at large scales in various contexts~\cite{Salopek:1990jq,Salopek:1990re,Prokopec:2019srf,Rigopoulos:2021nhv}. In particular, combining \Eq{eq:deltaH:deltaphi} with the Friedmann equation~\eqref{eq:Friedmann}
one obtains the Hamilton-Jacobi equation
\begin{equation}
\label{eqn:HJ}
    \left( \frac{\dd H}{\dd \phi} \right)^2 = \frac{3H^2}{2\Mp^2} - \frac{V\left(\phi\right)}{2\Mp^4} = \frac{\dot{\phi}^2}{4 \Mp^4} \,.
\end{equation}
This formula relies on one main assumption, namely that anisotropic degrees of freedom can be discarded in the momentum constraint to obtain \Eq{eq:MomentumConstraint:SalopekBond}. From the above consideration we see that this is equivalent to requiring that perturbations in the $(\phi, H)$ phase-plane are adiabatic. In following sections, we will show that this is justified on large scales only if a specific gauge is chosen or a dynamical attractor exists in the phase-plane. The latter is true in slow-roll inflation, but not during ultra-slow-roll inflation or other transient regimes.

\section{Perturbed spacetime}
\label{sec:pert}

Following the Arnowitt-Deser-Misner (ADM) formalism \cite{Arnowitt:1959ah}, we can foliate spacetime through spacelike hypersurfaces. We assume that there exists a homogeneous and isotropic FLRW background and we work in conformal time. We consider matter in the form of a scalar field $\phi$, whose canonical momentum is $\pi_\phi = a^2 \phi'$.
The time-reparameterisation invariance of the theory implies the presence of a background Hamiltonian constraint, namely the Friedmann equation~\eqref{eq:Friedmann}.

At first order in cosmological-perturbation theory, the line element of scalar perturbations in the ADM form can be expressed as~\cite{Arnowitt:1962hi,Langlois:1994ec,Malik:2008im}
\bea
\label{eq:metric:ADM}
	\dd s^2 &=& - a^2 \left(1+2 \frac{\delta N}{a} \right) \dd \eta^2 + 2 a^2 \delta_{ij} \delta N^{i} \dd \eta \,\dd x^j + \, a^2 \left( \delta_{ij} + \frac{\delta\gamma_{ij}}{a^2} \right) \dd x^i \dd x^j \,.
\eea
Here, $\delta N$ and $\delta N^i$ denote the perturbations of the lapse function and shift vector respectively, while $\delta \gamma_{ij}$ represents perturbations of the spatial metric. We highlight the fact that, as the FLRW background is homogeneous and isotropic, all spatial dependence appears at first order only in the perturbations, \eg $\phi(\eta,\vec{x})\rightarrow \phi(\eta) + \delta \phi\left(\eta, \vec{x}\right)$. We will abuse notation and denote the background variables as $\phi$, $\pi_\phi$, etc. In particular, we choose a background parameterisation such that the background shift vector vanishes. 

Working in Fourier space, the variance of the field is given by the sum over all modes
\bea
\langle \delta X^2\left(\eta,\vec{x}\right) \rangle = \int \frac{4\pi k^3}{(2\pi)^3}  \left|\delta X\left(\eta,k\right)\right|^2 \dd \ln k \,. \label{eq:Fourier}
\eea
The dimensionless power spectrum is thus given by
\begin{equation}
    {\cal P}_{X}(\eta,k) = \frac{k^3}{2\pi^2} \left| \delta X^2(\eta,k) \right| \,,
\end{equation}
from which we see that Fourier modes behaving as $\delta X(\eta,\vec{k}) \propto k^{-3/2}$ correspond to a scale-invariant power spectrum. 

Following \Refc{Artigas:2021zdk} we decompose spatial metric perturbations into trace and tracefree parts in Fourier space,
\bea
\delta\gamma_{ij}\left(\eta,\vec{k}\right) &=& M^1_{ij} \,\delta\gamma_1\left(\eta,\vec{k}\right) + M^2_{ij}\left(\vec{k}\right) \delta\gamma_2\left(\eta,\vec{k}\right) \,,
\eea
where we introduce the orthonormal basis tensors
\bea
M^1_{ij} = \frac{\delta_{ij}}{\sqrt{3}}\, ,  \qquad  \qquad M^2_{ij}\left(\vec{k}\right)= \sqrt{\frac{3}{2}} \left( \frac{k_i k_j}{k^2} - \frac{\delta_{ij}}{3} \right) ,
\eea
such that  $M^\alpha_{ij} M_\beta^{j \ell} = \delta^\alpha_\beta \delta_i^\ell$. The conformal perturbation, $\delta\gamma_1$, represents perturbations of the trace of the spatial metric and therefore describes isotropic perturbations, while $\delta\gamma_2$ describes the tracefree, hence anisotropic, perturbations. \App{app:lag} presents a translation between these perturbations and other variables commonly used in the literature.

The phase space is six-dimensional and is totally described by the following sets of canonical variables: $\left(\delta \gamma_1, \delta \pi_1 \right)$ and $\left(\delta \gamma_2, \delta \pi_2 \right)$ for metric perturbations and $\left(\delta \phi, \delta \pi_\phi \right)$ for scalar-field perturbations. We define the momenta as usual through the variation of the action with respect to their generalised coordinates, and they must obey the Poisson brackets
\bea
\left\{ \delta X^\mu\left(\eta_1,\vec{x}\right) , \delta \Pi_\nu \left(\eta_2 ,\vec{y}\right) \right\} = \delta^\mu_\nu\, \delta\left(\eta_1-\eta_2\right) \delta^{(3)}\left(\vec{x}-\vec{y}\right) \, ,
 \eea
where  $X^\mu=\left(\delta\gamma_1, \delta\gamma_2, \delta\phi\right)$ and $\Pi_\nu=\left(\delta\pi_1, \delta\pi_2, \delta\pi_\phi\right)$.

The perturbed lapse and shift, $\delta N$ and $\delta N^i$, play the role of Lagrange multipliers, which lead to the existence of the scalar and the momentum constraints, $\delta\mathcal{S}=0$ and $\delta{\mathcal{D}}_i= 0$,  respectively. Considering only scalar perturbations, we rewrite in Fourier space the shift and the momentum constraint as $\delta N^i(t, \vec{k}) = i k^i \delta N_1(t, \vec{k}) / k $ and $\delta{\mathcal{D}}_i(t, \vec{k})=i k_i \delta{\mathcal{D}}(t, \vec{k})$. The two linear constraints then read \cite{Artigas:2021zdk}
\bea
	\delta\mathcal{S}
 &=& 2\sqrt{3}a^2 H\,\delta\pi_1-\frac{a}{\sqrt{3}}\left[\frac{\pi_\phi^2}{a^6}-V\left(\phi\right) +{\Mp^2}\frac{k^2}{a^2}\right]\,\delta\gamma_1+\frac{\Mp^2}{\sqrt{6}}\frac{k^2}{a}\delta\gamma_2
\nonumber\\ 
&&	+\frac{\pi_\phi}{a^3}\,\delta\pi_\phi + a^3V_{,\phi}\delta\phi\,,
	\label{eq:scal} \\
	\delta{\mathcal{D}}
 &=& \pi_\phi\,\delta\phi -\frac{2\Mp^2}{\sqrt{3}}a H\left(\frac{1}{2}\delta\gamma_1-\sqrt{2}\delta\gamma_2\right)-\frac{2}{\sqrt{3}}a^2\left(\delta\pi_1+\sqrt{2}\delta\pi_2\right)\, . \label{eq:mom}
\eea

\subsection{Gradient expansion}

It is often helpful to perform an expansion of inhomogeneous perturbations in terms of spatial-gradient terms~\cite{Salopek:1990jq,Shibata:1999zs,Comer:1994np}. We can expand a particular solution for any perturbation variable, $\delta X_{\mathrm{p}}$, in Fourier space as
\bea
\label{def:gradientexpansion_particular}
\delta X_{\mathrm{p}}\left(\eta,k\right) = f_{\mathrm{p}}(k) \sum_{n=0}^\infty k^{2n} \delta X^{(n)}(\eta)\,.
\eea
The Einstein equations then only involve terms of order $k^{2m}$ in this gradient expansion, where $m$ is a non-negative integer. For example, the scalar constraint equation \eqref{eq:scal} relates terms at order $k^{2n}$ and $k^{2(n+1)}$ in the gradient expansion \eqref{def:gradientexpansion_particular}, while the momentum constraint \eqref{eq:mom} only relates terms of the same order $k^{2n}$. Higher-order terms in the gradient expansion are obtained by iteratively solving the equations of motion for $\delta X^{(n+1)}(\eta)$ in terms of $\delta X^{(n)}(\eta)$ \cite{Comer:1994np}.

The general solution of this second-order system is thus
\bea
\label{def:gradientexpansion}
\delta X\left(\eta,k\right) = \sum_{n=0}^\infty 
f_+(k) k^{2n}
\delta X_+^{(n)}(\eta) + 
f_-(k) k^{2n}
\delta X_-^{(n)}(\eta)\,,
\eea
where $\delta X_+^{(0)}(\eta)$ and $\delta X_-^{(0)}(\eta)$ are two linearly independent solutions in the long-wavelength limit where we set $k\to0$ in the Klein-Gordon and Einstein equations. Their overall $k$-dependence is given by the functions $f_+(k)$ and $f_-(k)$ which are   determined by initial conditions.
We will investigate examples corresponding to simple choices for the initial state of field perturbations during inflation in \Sec{sec:appli} which give simple power laws, $f_\pm(k)\propto k^{p_\pm}$, where $p_\pm=-3/2$ corresponds to a scale-invariant power spectrum.

Similarly, we expand the constraints \eqref{eq:scal} and \eqref{eq:mom} as follows:
\bea
\delta\mathcal{S}\left(\eta,\vec{k}\right) = \sum_{n=0}^\infty 
f_+(k)
k^{2n} 
\delta\mathcal{S}_+^{(n)}(\eta) + 
f_-(k)
k^{2n} 
\delta\mathcal{S}_-^{(n)}(\eta)\,, \label{eq:Sn} \\
\delta{\mathcal{D}}\left(\eta,\vec{k}\right) = \sum_{n=0}^\infty 
f_+(k)
k^{2n} 
\delta\mathcal{D}_+^{(n)}(\eta) + 
f_-(k)
k^{2n} 
\delta\mathcal{D}_-^{(n)}(\eta)\,. \label{eq:Dn}
\eea
In the large-scale/long-wavelength limit we keep only the leading term in the gradient expansion~\eqref{def:gradientexpansion}. 
In particular from \Eqs{eq:scal} and~\eqref{eq:mom} we have,
\bea
	\delta{\mathcal{S}}^{(0)}
 &=& 2\sqrt{3}a^2 H\,\delta\pi_1^{(0)}-\frac{a}{\sqrt{3}}\left[\frac{\pi_\phi^2}{a^6}-V\left(\phi\right) 
 \right]\,\delta\gamma_1^{(0)}
+\frac{\pi_\phi}{a^3}\,\delta\pi_\phi^{(0)} + a^3V_{,\phi}\delta\phi^{(0)}\,,
	\label{eq:scal0} \\
	\delta{\mathcal{D}}^{(0)}
 &=& \pi_\phi\,\delta\phi^{(0)} -\frac{2\Mp^2}{\sqrt{3}}a H\left(\frac{1}{2}\delta\gamma_1^{(0)}-\sqrt{2}\delta\gamma_2^{(0)}\right)-\frac{2}{\sqrt{3}}a^2\left(\delta\pi_1^{(0)}+\sqrt{2}\delta\pi_2^{(0)}\right)\,. \label{eq:mom0}
\eea

In general the perturbation variables, such as the scalar-field perturbation, $\delta\phi$, or the metric perturbations, $\delta\gamma_i$, are gauge dependent, and thus the terms in the gradient expansion will be gauge dependent. Hence we often prefer to work with gauge-invariant variables~\cite{Bardeen:1980kt}. For example, the gauge-invariant Bardeen potential, $\Psi$, describes the spatial metric perturbation $\delta\gamma_1^{\text{(N)}}$ in the Newtonian (or longitudinal) gauge in which $\delta\gamma_2^{\text{(N)}}=\delta\pi_2^{\text{(N)}}=0$~\cite{Bardeen:1980kt,Mukhanov:1990me}, while the gauge-invariant Mukhanov-Sasaki~\cite{Sasaki:1986hm,Mukhanov:1988jd} variable, $u$, describes the scalar-field perturbation, $a\delta\phi^{\text{(SF)}}$, in the spatially-flat gauge in which $\delta\gamma_1^{\text{(SF)}}=\delta\gamma_2^{\text{(SF)}}=0$~\cite{Kodama:1984ziu,Malik:2008im}. The same physical solution will have different gradient expansions in terms of different gauge-invariant variables. The scale-dependence of physical observables should be gauge invariant, but gauge-dependent variables may have different scale dependence in different gauges, hence arise at different orders in the gradient expansion.

\subsection{Isotropic-anisotropic splitting}

To facilitate the comparison with the usual separate-universe approach, we split the constraints \eqref{eq:scal} and \eqref{eq:mom} into an isotropic part, denoted by an overbar, and an anisotropic part, denoted by a tilde
\bea
\delta \mathcal{S} &\equiv & \ovl{\delta \mathcal{S}} + \widetilde{\delta \mathcal{S}} \,, \\
\delta \mathcal{D} &\equiv & \ovl{\delta \mathcal{D}} + \widetilde{\delta \mathcal{D}} \,. 
\eea
The isotropic parts depend only on $\left(\delta\phi, \delta\gamma_1,\delta\pi_\phi, \delta\pi_1\right)$ and we have
\bea
	\ovl{\delta \mathcal{S}}
 &=& 2\sqrt{3}a^2 H\,\delta\pi_1-\frac{a}{\sqrt{3}}\left[\frac{\pi_\phi^2}{a^6}-V\left(\phi\right) +{\Mp^2}\frac{k^2}{a^2}\right]\,\delta\gamma_1
\nonumber\\ 
&&	+\frac{\pi_\phi}{a^3}\,\delta\pi_\phi + a^3V_{,\phi}\delta\phi\,,
	\label{eq:scalbar} \\
	\ovl{\delta \mathcal{D}}
 &=& \pi_\phi\,\delta\phi -\frac{\Mp^2}{\sqrt{3}}a H\delta\gamma_1-\frac{2}{\sqrt{3}}a^2\delta\pi_1\,, \label{eq:mombar}
\eea
while the anisotropic part depends on $\left(\delta\gamma_2,\delta\pi_2\right)$
\bea
	\widetilde{\delta\mathcal{S}}
 &=& \frac{\Mp^2}{\sqrt{6}}\frac{k^2}{a}\delta\gamma_2
\,,
	\label{eq:scaltilde} \\
	\widetilde{\delta{\mathcal{D}}}
 &=& \frac{2\sqrt{2}}{\sqrt{3}}\left( \Mp^2 a H\delta\gamma_2
 -a^2\delta\pi_2 \right)\,. \label{eq:momtilde}
\eea
It is instructive to write the isotropic parts of the constraints in terms of the perturbed expansion rate
\bea
 \delta H = -\frac{H}{2\sqrt{3}a^2} \delta \gamma_1 - \frac{1}{\sqrt{3}\Mp^2 a} \delta \pi_1 \,,
 \label{deltaH}
\eea
and the perturbed energy density
\bea
a^3 \delta\rho = \frac{\pi_\phi}{a^3} \delta\pi_\phi - \frac{\sqrt{3}}{2} \frac{\pi_\phi^2}{a^5} \delta\gamma_1 + a^3 V_{,\phi} \delta\phi \, . \label{deltaRho}
\eea 
Thus we have the large-scale limit of the isotropic parts of the constraints \eqref{eq:scalbar} and \eqref{eq:mombar}
\bea
\label{iso-energy-H}
	\ovl{\delta \mathcal{S}}^{(0)}
 &=& a^3 \left( \delta\rho^{(0)} - 6 \Mp^2 H \delta H^{(0)} \right)
 \,, \\
 \label{iso-mtm-H}
	\ovl{\delta \mathcal{D}}^{(0)}
 &=& a^3 \left(\dot{\phi} \delta\phi^{(0)} + 2 \Mp^2 \delta H^{(0)} \right) \,. 
\eea
While the anisotropic part of the energy constraint \eqref{eq:scaltilde} automatically vanishes in this large-scale limit, 
\bea
\widetilde{\delta\mathcal{S}}^{(0)}=0 \,,
\eea
the anisotropic part of the momentum constraint \eqref{eq:momtilde} does not in general vanish on large scales, 
\bea
	\widetilde{\delta{\mathcal{D}}}^{(0)}
 &=& \frac{2\sqrt{2}}{\sqrt{3}}\left( \Mp^2 a H\delta\gamma_2^{(0)}
 -a^2\delta\pi_2^{(0)} \right). \label{eq:momtilde0}
\eea

The energy constraint~\eqref{eq:scal} thus reduces to the isotropic constraint~\eqref{iso-energy-H} on large scales, which relates the perturbed expansion rate to the perturbed energy density as would be obtained from directly perturbing the background Friedmann equation~\eqref{eq:Friedmann}, consistent with the separate-universe approach. However the momentum constraint~\eqref{eq:scal} cannot necessarily be reduced to its isotropic part~\eqref{iso-mtm-H} on large scales.

We see that setting the isotropic part of the momentum constraint \eqref{iso-mtm-H} to zero on large scales is required to impose the relation \eqref{eq:deltaH:deltaphi}. 
We will refer to this as the Hamilton-Jacobi constraint. The Hamilton-Jacobi approach \cite{Salopek:1990jq, Salopek:1990re, Prokopec:2019srf, Rigopoulos:2021nhv} therefore implicitly discards the anisotropic part of the constraints at large scales~\cite{Tomberg:2022mkt}. We will now show that the presence of non-adiabatic pressure perturbations on large scales can invalidate the use of the Hamilton-Jacobi approach. This is reminiscent of the fact that higher orders of the gradient expansion can become relevant in models beyond slow roll~\cite{Leach:2001zf}. 

\section{The non-adiabatic pressure perturbation}
\label{sec:dpnad}

The perturbed energy density is given in \Eq{deltaRho} while the perturbed pressure is obtained by replacing $V_{,\phi}\rightarrow -V_{,\phi}$ in this equation. 
In terms of field and metric perturbations, an explicit computation thus leads us to the following equation for the non-adiabatic pressure \eqref{def:Pnad}:
\bea
\delta \Pnad = \left(-2 V_{,\phi} - \frac{2}{3} \frac{a^3 {V_{,\phi}}^2}{\pi_\phi H} \right) \delta\phi + \left( \frac{1}{\sqrt{3}} \frac{V_{,\phi} \pi_\phi}{a^5 H}  \right) \delta\gamma_1 - \left( \frac{2}{3} \frac{ V_{,\phi}}{a^3 H}  \right)  \delta\pi_\phi\,, \label{eq:dPnad}
\eea
where the adiabatic sound speed for the scalar field is
\bea
\cs^2 = \frac{\dot{P}}{\dot{\rho}} = 1 + \frac{2}{3} \frac{a^3 V_{,\phi}}{\pi_\phi H}\,.
\eea

We can then rewrite the scalar constraint \eqref{eq:scal} in terms of the non-adiabatic pressure perturbation \eqref{eq:dPnad} and the isotropic part of the momentum constraint \eqref{eq:mombar} to obtain \cite{Naruko:2012fe}
\bea
-\frac{1}{3 H} \delta\mathcal{S} = \ovl{\delta\mathcal{D}} + \frac{\pi_\phi}{2V_{,\phi}} \delta \Pnad + \frac{\Mp^2 k^2}{3\sqrt{3}} \frac{\delta\gamma_1}{a H} - \frac{\Mp^2 k^2}{3\sqrt{6}} \frac{\delta\gamma_2}{a H}\,. \label{D+dPnad+k}
\eea
The scalar constraint $\delta \mathcal{S}=0$ can be solved at an initial time, and is then automatically conserved over time on the surface of constraints.
At leading order in the gradient expansion we drop the terms of order $k^2$ and higher in \Eq{D+dPnad+k} and we have
\bea
\ovl{\delta\mathcal{D}}^{(0)} = - \frac{\pi_\phi}{2V_{,\phi}} \delta \Pnad^{(0)}  \,. 
\label{Dbar0=dPnad0}
\eea

We see that imposing the Hamilton-Jacobi constraint, $\ovl{\delta\mathcal{D}}^{(0)}=0$, requires the non-adiabatic pressure perturbation \eqref{eq:dPnad} to vanish in the large-scale limit. In this case we can recover the results using the Hamilton-Jacobi approach, see, \eg,  Eq.~(2.11) of \cite{Salopek:1990jq} or Eq.~(2.8) of \cite{Rigopoulos:2021nhv}. However, since $\delta\Pnad$ is gauge invariant, setting $\delta\Pnad^{(0)}=0$ corresponds to a physical restriction on the solutions that can be obtained in this large-scale limit. 

For example, when working in the spatially-flat gauge, where $\delta\gamma_1^{(\text{SF})}=\delta\gamma_2^{(\text{SF})}=0$, one obtains from \Eq{D+dPnad+k}
\bea
\ovl{\delta\mathcal{D}}^{(\text{SF})} = - \frac{\pi_\phi}{2V_{,\phi}} \delta \Pnad\,, \label{Dbar=dPnad}
\eea
on all scales.
If we wish to describe the effect of non-adiabatic field fluctuations in scenarios beyond slow-roll inflation, then we should not set
$\ovl{\delta\mathcal{D}}^{(\text{SF})}$ to zero in the spatially-flat gauge. In the next section, we will compute $\delta\Pnad$ and $\ovl{\delta\mathcal{D}}^{(\text{SF})}$ in terms of field and metric perturbations in the spatially-flat gauge to explicitly show that they do not vanish on large scales for the Bunch-Davies vacuum state in ultra-slow roll.

Conversely, when working in the Newtonian gauge where $\delta\gamma_2^{\text{(N)}} =  \delta\pi_2^{\text{(N)}} =0$~\cite{Bardeen:1980kt,Mukhanov:1990me} the anisotropic part of the momentum constraint~\eqref{eq:momtilde} vanishes by construction, and then the isotropic part of the momentum constraint is automatically satisfied, $\ovl{\delta\mathcal{D}}^{\text{(N)}}=0$. As a consequence, the Hamilton-Jacobi constraint \eqref{eq:deltaH:deltaphi} is valid in the Newtonian gauge. In this case the energy constraint~\eqref{D+dPnad+k} reduces to
\bea
\label{eq:dPnadN}
\frac{\pi_\phi}{2V_{,\phi}}  \dPnad = - \frac{\Mp^2 k^2}{3\sqrt{3}a H} \delta\gamma_1^{\text{(N)}} \,.
\eea
As a consequence, the non-adiabatic pressure perturbation appears only at first order in a gradient expansion in the Newtonian gauge \cite{Seto:1999jc,Leach:2001zf,Takamizu:2010xy,Naruko:2012fe,Artigas:2024xhc}.

Thus the validity of the Hamilton-Jacobi equation in the large-scale limit may depend upon the choice of perturbation variables, or equivalently the choice of gauge used to describe field and metric perturbations. If one studies the evolution of scalar field fluctuations in the spatially-flat gauge one can include non-adiabatic field fluctuations and study the evolution of the comoving curvature perturbation, $\dot{\cal R}^{(0)}\neq0$, in the large-scale limit. However if one works in the Newtonian gauge the field perturbations are constrained to be adiabatic, and thus $\dot{\cal R}^{(0)}=0$, in the large-scale limit, and the time-dependence of ${\cal R}$ can only arise at higher order in the gradient expansion in that gauge.

\section{Applications}
\label{sec:appli}

In this section we will present the gradient expansion for field perturbations in the spatially-flat gauge during slow-roll and ultra-slow-roll inflation. We will consider solutions of the Mukhanov-Sasaki equation~\cite{Sasaki:1986hm,Mukhanov:1988jd}
\begin{align}
u''+\left(k^2-\frac{z''}{z}\right)u =0\, , \label{eq:ms}
\end{align}
where we recall that $z=a\dot{\phi}/H=\sigma a\sqrt{2\epsilon_1}\Mp$, with $\epsilon_1:=-\dot{H}/H^2$ the first slow-roll parameter, and we define $\sigma = \text{sign}(\dot\phi)$. The Mukhanov-Sasaki variable 
\begin{align} 
\label{def:u}
u:= a\delta\phi -\frac{\pi_\phi}{2\sqrt{3} H a^{4}}\left(\delta\gamma_1-\frac{1}{\sqrt{2}}\delta\gamma_2\right)
\end{align}
is gauge invariant.
In the spatially-flat gauge, $u$ is directly proportional to the field perturbation, $u=a\delta\phi^{(\text{SF})}$, and the perturbed scalar-field momentum becomes \cite{Artigas:2021zdk}
\bea
\delta\pi_\phi^{(\text{SF})} &=& a^2 \left(\delta\phi^{(\text{SF})\prime} - a H \epsilon_1 \delta\phi^{(\text{SF})} \right) \,. \label{eq:dPiPhi}
\eea
In the remainder of this section we will work in the spatially-flat gauge and drop the (SF) superscript.
In this gauge, the non-adiabatic-pressure perturbation~\eqref{eq:dPnad} reduces to
\begin{align}
  \delta \Pnad =   \sigma \frac{\Mp H}{3 a^3} \sqrt{2\epsilon_1} \left(6 + \epsilon_2 - 2\epsilon_1 \right) \left[ \delta\pi_\phi - \frac{a^3H}{2}\left(\epsilon_2 - 2\epsilon_1 \right) \delta\phi \right] \,,
 \label{eq:dPnad_Epsilon}
\end{align}
where the background factors have been expressed in terms of slow-roll parameters, see \App{App:SRParam}.

The initial values of $\delta\phi$ and $\delta\pi_\phi$ (or equivalently $\delta\phi$ and $\delta\phi'$) can be set independently, and the constraints \eqref{eq:scal} and \eqref{eq:mom} then determine the remaining perturbations $\delta\pi_1$ and $\delta\pi_2$, remembering that $\delta\gamma_1$ and $\delta\gamma_2$ have already been set to zero in this gauge.

\subsection{Slow roll} \label{sec:SR-CPT}

First we consider the case of slow-roll inflation, where $\epsilon_1,|\epsilon_2| \ll 1$ and expand the solutions up to next-to-leading order in the slow-roll parameters.
By Taylor expanding  the slow-roll parameters about the time $\eta_*=-1/k$ of Hubble-radius exit, one finds
\bea
a &\simeq& - \frac{1}{H_* \eta} \left[1 + \epsilon_{1*} - \epsilon_{1*} \ln\left(\frac{\eta}{\eta_*}\right)\right], \label{eq:a} \\
H &\simeq& H_* \left[1 + \epsilon_{1*}  \ln\left(\frac{\eta}{\eta_*}\right)\right], \label{eq:H} 
\eea
see \App{App:SRExp}. 
Slow-roll inflation is known to exhibit an attractor solution meaning that $\phi'\to\phi'\left(\phi\right)$ in the background phase space and, as argued in \Sec{sec:intro}, we expect the non-adiabatic pressure \eqref{eq:dPnad_Epsilon} to vanish in the large-scale limit. We will show that this is indeed the case.

To determine the behaviour of $\delta \phi$, we solve the Mukhanov-Sasaki equation~\eqref{eq:ms}, in the spatially-flat gauge, setting the initial state at $k\eta\to-\infty$ in the Bunch-Davies vacuum state \cite{Stewart:1993bc,Pattison:2019hef},
\begin{align}
\label{eq:Qsr}
u = a\delta\phi = \frac{\sqrt{\pi}}{2}(-\eta)^{1/2}H_{\nu}^{(2)}(-k\eta)\,, 
\end{align}
where $H_{\nu}^{(2)}$ is the Hankel function of the second kind and 
\begin{align}
\label{def:nu}
\nu \simeq \frac32 + \epsilon_{1*}+\frac{\epsilon_{2*}}{2}  \,. 
\end{align}
In the large-scale limit, $\vert k\eta\vert\ll 1$, a gradient expansion for $H_{\nu}^{(2)}$ gives
\begin{align}
\label{eq:Hnu}
H_{\nu}^{(2)}(-k\eta) =  \frac{i \Gamma{(\nu)}}{\pi}\left(\frac{2}{-k\eta}\right)^{\nu} \left[ 1+ \frac{(k \eta)^2}{4 (\nu - 1)} + \mathcal{O}\left(k\eta\right)^4\right] - \frac{i e^{i\pi\nu}\left(-k\eta\right)^{\nu}}{2^\nu \Gamma\left(1+\nu\right)\sin(\pi\nu)}\left[ 1+\mathcal{O}\left(k\eta\right)^2\right].
\end{align}
Combining \Eqs{eq:Qsr}, \eqref{eq:Hnu} and \eqref{eq:a}, the scalar-field perturbations in slow roll, where $\nu\simeq 3/2$, are thus given by the gradient expansion
\bea
\label{eq:dphsr}
\delta \phi &=& f_+(k) \left[\delta \phi_+^{(0)}(\eta)+k^2\delta \phi_+^{(1)}(\eta)+\mathcal{O}(k^4)\right]+\mathcal{O}(k^\nu\eta^{\nu+3/2})
\eea 
where 
\bea 
f_+(k) = i H_* \frac{\Gamma(\nu)}{2\sqrt{\pi}}\left(\frac{k}{2}\right)^{-\nu} \,,
\eea 
together with
\bea
\delta \phi_+^{(0)}(\eta)=(-\eta)^{3/2-\nu}\left[1-\epsilon_{1*}+\epsilon_{1*}\ln\left(\frac{\eta}{\eta_*}\right)\right] \,,
\quad
\delta \phi_+^{(1)}(\eta)=\frac{\eta^2}{4(\nu-1)}\delta \phi_+^{(0)}
\,.
\eea
On the one hand we see that the leading term, proportional to $f_+(k)\propto k^{-\nu}$, in \Eq{eq:dphsr} gives rise to an approximately scale-invariant growing mode in the long-wavelength limit as expected, since from \Eq{def:nu} we have $\nu\simeq3/2$ in slow-roll inflation. On the other hand, the term proportional to $k^{\nu}$ in \Eq{eq:dphsr} corresponds to the leading-order term in a gradient expansion for the decaying mode\footnote{When writing $H_\nu^{(2)}$ in terms of Bessel functions $J_\nu$ and $Y_\nu$, this terms comes from the series expansion of the regular Bessel function $J_\nu$.} and can be neglected with respect to the $k^2f_+(k)$-term for $\nu>1$ on super-Hubble scales.

From the definition of $\delta\pi_\phi$ in terms of $\delta\phi'$, \Eq{eq:dPiPhi}, we obtain
\bea
\delta\pi_\phi = f_+(k) \left[\delta\pi_+^{(0)}(\eta)+k^2\delta\pi_+^{(1)}(\eta)+\mathcal{O}(k^4)\right]+\mathcal{O}\left(k^\nu \eta^{\nu-3/2}\right)
\eea 
where
\bea
\delta\pi_+^{(0)} &=& \frac{(-\eta)^{-3/2-\nu}}{H_*^2}\left(\frac{\epsilon_{2*}}{2}-\epsilon_{1*}\right) ,
\\
\delta\pi_+^{(1)} &=& \frac{(-\eta)^{1/2-\nu}}{2(1-\nu)H_*^2}\left[1+\frac{3}{2}\epsilon_{1*}-\frac{\epsilon_{2*}}{4}-\epsilon_{1*} \ln\left(\frac{\eta}{\eta_*}\right)\right]
.
\eea 
One notices that, at leading order in gradient expansion, $\delta\pi_\phi$ is also approximately scale invariant but is of order $\epsilon_i$ in the slow-roll expansion. 

From the above expressions we note that the combination $2\delta\pi_\phi-a^3 H (\epsilon_2-2\epsilon_1)\delta\phi$, appearing in \Eq{eq:dPnad_Epsilon} for the non-adiabatic pressure, vanishes at zero-th order in the gradient expansion, hence 
\bea
\label{eq:deltaPnad:SR:0}
\delta P_{\mathrm{nad}+}^{(0)} = 0\, .
\eea
Up to first order in the gradient expansion, one finds
\bea
\label{eq:dPnad_SRbis}
\delta \Pnad =  -2\sigma f_+(k) \Mp H_*^2\sqrt{2\epsilon_{1*}} \left[1+\mathcal{O}(\epsilon_i)\right] (k\eta)^2 .
\eea 
From \Eq{Dbar=dPnad} for $\ovl{\delta\mathcal{D}}$ in the spatially-flat gauge we conclude that requiring the isotropic part of the momentum constraint to vanish at zero-th order in the gradient expansion (\ie, $\ovl{\delta\mathcal{D}}^{(0)}=0$) is valid for the Bunch-Davies vacuum state in the slow-roll approximation. Indeed, the fact that the momentum constraint is gradient suppressed was already shown for the case of slow-roll evolution  \cite{Sugiyama_2013} or more generally in the presence of any background attractor \cite{Garriga:2016poh}.

\subsection{Ultra-slow roll}

In a USR phase, the potential is assumed to be almost flat and the Klein-Gordon equation~\eqref{eq:KG} reduces to
\bea
	\phi^{\prime \prime} + 2 a H \phi^{\prime} \simeq 0 \,.
\eea
This equation determines the rate at which $\vert{\phi^{\prime}}\vert$ decreases in the case where  $V_{,\phi}=0$, which gives $\phi^{\prime}\propto a^{-2}$. In this scenario, the second slow-roll parameter $\epsilon_2=-6$ and, since $\epsilon_1\propto a^{-6}$ decreases rapidly, we will neglect terms of order $\mathcal{O}\left(\epsilon_1\right)$. 
At this order $H$ is constant and the scale factor reduces to the following relation (see \App{App:USRExp} for details):
\bea
a = - \frac{1}{H\eta}\,, \label{eq:aUSR}
\eea 
which yields $z''/z = 2 a^2 H^2$. When choosing the initial vacuum state to be that of Bunch-Davies, the solution to the Mukhanov-Sasaki equation \eqref{eq:ms} gives the field perturbations in the spatially-flat gauge \cite{Pattison:2019hef}
\bea
	\delta \phi =- \frac{H \eta}{\sqrt{2k}} \left(1-\frac{i}{k\eta} \right) e^{-i k\eta} \,. \label{eq:USRBD}
\eea
Expanding the exponential in \eqref{def:gradientexpansion} up to order $\mathcal{O}\left(k\eta\right)^2$, we obtain the first two terms in the gradient expansion
\bea
\delta\phi &=& f_+(k) \left[1+\frac{(k\eta)^2 }{2}  \right] + {\cal O}(k\eta)^{3/2}\,,
\eea
where we pull out the $k$-dependent factor
\bea
    f_+(k)&=&\frac{i H}{\sqrt{2}} k^{-3/2} \,.
\eea
Taking the conformal-time derivative and substituting this into the momentum~\eqref{eq:dPiPhi}, we find the leading-order terms in gradient expansion of $\delta\pi_\phi$,
\bea
\label{eq:dPi-USR}
\delta\pi_\phi &=& f_+(k) \left( \delta\pi_\phi^{(0)} + k^2 \delta\pi_\phi^{(1) }  \right) + {\cal O}(k\eta)^{3/2}\,,
\eea
where
\bea
\delta\pi_\phi^{(0)} = 0\,, \quad \delta\pi_\phi^{(1)} = \frac{1 }{ H^2\eta} \,.
\eea

In USR, the non-adiabatic pressure is small since it is proportional to $V_{,\phi} \simeq 0$. Keeping explicit the presence of the potential derivative, \Eq{eq:dPnad} then reduces to
\bea
\label{eq:dPnadUSR}
\delta \Pnad &=& -2V_{,\phi} \left[\delta\phi + \frac{H^2 \left(-\eta\right)^3}{3} \delta\pi_\phi \right] .
\eea
The leading-order terms in the gradient expansion are thus
\bea
\delta \Pnad &=& f_+(k) \left( \delta P_{\rm nad+}^{(0)} + k^2 \delta P_{\rm nad+}^{(1)} \right) + {\cal O}(k\eta)^{3/2} \,,
\eea
where
\bea
\delta P_{\rm nad+}^{(0)} = -2 V_{,\phi}  \quad , \quad  \delta P_{\rm nad+}^{(1)} = -\eta^2 V_{,\phi}/3  \,.\label{eq:dPnad-USR}
\eea
At leading order in the gradient expansion, $\dPnad$ is non-zero for $V_{,\phi}\neq0$ and is scale invariant, $\dPnad\propto k^{-3/2}$. 

A similar conclusion holds at the level of the isotropic part of the momentum constraint~\eqref{Dbar=dPnad}, given by
\begin{align}
\overline{\delta\mathcal{D}} = - \sigma \Mp\frac{\sqrt{2 \epsilon_1}}{H^2 \eta^3} f_+(k) \left( 1 + \frac{k^2\eta^2}{6}  \right) + {\cal O}(k\eta)^{3/2} \,.
\end{align}
While the Hubble rate is constant at leading order in $\epsilon_1$, the first slow-roll parameter decays as $\epsilon_1\propto \eta^6$, so $\sqrt{2 \epsilon_1}/H^2\eta^3$ remains constant, and we see that $\ovl{\delta\mathcal{D}}^{(0)}$ is conserved \cite{Artigas:2021zdk}, but non zero. 
Although the non-adiabatic pressure perturbation \eqref{eq:dPnadUSR} is small (on all scales) because of the overall factor $V_{,\phi}$, this factor does not appear in the isotropic part of the momentum constraint. 
We conclude that $\ovl{\delta \mathcal{D}}$ is non-vanishing at large scales in the spatially-flat gauge for the Bunch-Davies vacuum state in USR.

\subsection{Gradient expansion of the anisotropic sector}
As a final remark, let us comment on the behaviour of the anisotropic sector. Since the complete momentum constraint is $\overline{\delta \mathcal{D}}=-\widetilde{\delta \mathcal{D}}$, the anisotropic sector can be directly related to the non-adiabtic pressure perturbation. In the spatially-flat gauge, this is simply
\bea
\label{eq:dPnadSF}
\delta\pi_2 = - \sqrt{\frac{3}{2}} \frac{\pi_\phi}{4 a^2 V_{,\phi}} \dPnad\,.
\eea
In the slow-roll approximation, \Eq{eq:deltaPnad:SR:0} shows that at leading order in the gradient expansion the anisotropic sector can be safely neglected, $\delta\pi_2^{(0)}=0$. In the USR case, however, \Eq{eq:dPnad-USR} yields
\bea
\delta\pi_2^{(0)} &=& - \frac{\sqrt{3\epsilon_1}}{2 \eta} \sigma \Mp  \,,
\eea
when neglecting terms $\mathcal{O}\left(\epsilon_1\right)$. Ignoring the anisotropic perturbations is therefore not justified beyond slow roll and leads to inconsistencies.

Finally, let us comment on the behaviour of the traceless part of the extrinsic curvature $A^i_j$. The latter can be related to $\delta\pi_2$ in the spatially-flat gauge as follows (see Appendix~\ref{app:lag}):
\bea
\delta A^i_{j} = - \frac{k^i k_j}{k^2} \frac{\sqrt{6}}{a \Mp^2} \delta\pi_2  \,, \qquad \forall \, i\neq j \,.
\eea
By solving the equation of motion for $\delta\pi_2$ in the spatially-flat gauge, see Eqs.~(3.32), (3.36) and (3.39) in \cite{Artigas:2021zdk}, one finds that at zero-th order in the gradient expansion $\delta\pi_2^{(0)} \propto a^{-2}$ in conformal time, meaning that
\bea
\delta A^{i(0)}_j \propto a^{-3} \,. 
\eea
This is consistent with the analysis by Salopek and Bond~\cite{Salopek:1990jq} who went on to argue that one could thus neglect contributions from the anisotropic sector on large scales and in particular, that one could impose the Hamilton-Jacobi constraint~\eqref{iso-mtm-H}. We have shown that this is equivalent to neglecting the non-adiabatic pressure on large scales, \Eq{Dbar0=dPnad0}.
However, using \eqref{eq:dPnadSF}, one finds that in general
\bea
\dPnad^{(0)} \propto 
\frac{V_{,\phi}}{\pi_\phi}  
\propto
\frac{V_{,\phi}}{a^3 H \sqrt{\epsilon_1}} \,.
\eea
While in slow roll we have $\sqrt{\epsilon_1}\propto V_{,\phi}/H^2$ and thus $\dPnad^{(0)}\propto Ha^{-3}$ corresponds to a decaying mode on large scales, in ultra-slow roll we have $\sqrt{\epsilon_1}\propto H^{-1}a^{-3}$ and thus $\dPnad^{(0)}\propto V_{,\phi}$ corresponds to the growing mode, \Eq{eq:dPnad-USR}.

\section{Conclusions}
\label{sec:conc}

The separate-universe approach~\cite{Wands:2000dp, Lyth:2003im, Rigopoulos:2003ak, Lyth:2004gb} is a powerful approach to treat perturbations non-linearly. This uses the equations of motion for a {\it homogeneous} cosmology to follow the local evolution in each patch of an {\it inhomogeneous} universe smoothed on a suitably large scale (usually the Hubble scale or larger). Such an approach considerably simplifies the non-linear evolution requiring only the solution of ordinary differential equations for the local time evolution, rather than non-linear partial-differential equations. This neglects spatial gradients, and thus we expect the separate-universe approach to correspond to taking the zero-th order limit of a gradient expansion in cosmological-perturbation theory~\cite{Salopek:1990jq}. 

The Hamilton-Jacobi approach~\cite{Salopek:1990jq, Salopek:1990re,Rigopoulos:2003ak} further simplifies calculations within each separate-universe patch by reducing the phase space to a one-dimensional trajectory, $H(\phi)$. This approach uses the scalar and momentum constraints on large scales, to obtain the Hamilton-Jacobi constraint \eqref{eq:deltaH:deltaphi} \cite{Salopek:1990jq,Salopek:1990re,Prokopec:2019srf,Rigopoulos:2021nhv}. However, the solution of the constraints is obtained after having discarded the anisotropic part of the momentum constraint \eqref{eq:momtilde}. The Hamilton-Jacobi approach therefore requires the vanishing of the isotropic and anisotropic parts of the momentum constraint separately.

In the Newtonian gauge, anisotropic metric perturbations are set to zero ($\delta\gamma_2^{\text{(N)}} =  \delta\pi_2^{\text{(N)}} =0$~\cite{Bardeen:1980kt,Mukhanov:1990me}) and thus when we drop spatial gradients at zero-th order in the gradient expansion we automatically recover the dynamics of a homogeneous and isotropic FLRW cosmology \cite{Artigas:2021zdk}. The anisotropic part of the momentum constraint~\eqref{eq:momtilde} vanishes by construction, and therefore the isotropic momentum constraint at leading order in a gradient expansion~\eqref{iso-mtm-H} imposes the Hamilton-Jacobi constraint \eqref{eq:deltaH:deltaphi}. As a consequence we can only describe adiabatic scalar-field perturbations, along the background trajectory, in the large-scale limit. This approach is thus restricted to describing a 1D phase space, $H(\phi)$. In this gauge the non-adiabatic pressure, $\dPnad$ in \eqref{eq:dPnadN}, only appears at higher order in the gradient expansion and capturing the full evolution beyond slow roll requires a solution at higher order in the gradient expansion~\cite{Takamizu:2010xy, Takamizu:2010je, Naruko:2012fe,Artigas:2024xhc,Launay:2024qsm}.

Conversely, if we work in the spatially-flat gauge ($\delta\gamma_1^{\text{(SF)}}=\delta\gamma_2^{\text{(SF)}}=0$~\cite{Kodama:1984ziu,Malik:2008im}) we can recover the full 2D phase space, $H(\phi,\dot\phi)$, in the large-scale limit where we neglect the spatial gradients. In this paper we have shown that, in linear-perturbation theory, the anisotropic part of the momentum constraint~\eqref{eq:momtilde} may remain finite in the large-scale limit, and indeed we have shown that this is necessary in order to describe non-adiabatic scalar-field perturbations \eqref{eq:dPnadSF}. However we cannot then apply the Hamilton-Jacobi approach 
in the spatially-flat gauge on large scales.

One might worry that the local dynamics cannot be described by a FLRW cosmology when anisotropic perturbations remain finite on large scales. However the separate-universe approach uses only the homogeneous-scalar-field evolution equation and the scalar (Friedmann) constraint to describe the local evolution of the scalar field. The anisotropy appears solely in the momentum constraint~\eqref{eq:mom}, and plays no role in the scalar constraint~\eqref{eq:scal} (since $\delta\gamma_2^{\text{(SF)}}=0$ in the spatially-flat gauge). Thus the scalar-field perturbations in this large-scale limit can be identified with evolution of a homogeneous scalar field in an isotropic FLRW spacetime, and the momentum constraint is only required to relate the local frame in a separate-universe patch to the global background chart in perturbation theory~\cite{Pattison:2019hef,Tanaka:2021dww}. 

The isotropic part of the momentum constraint in the spatially-flat gauge is proportional to the non-adiabatic pressure perturbation. By setting it to zero, the Hamilton-Jacobi approach therefore requires scalar-field fluctuations to follow the background phase-space trajectory. The isotropic part of the momentum constraint has been shown to be gradient-suppressed in the presence of a slow-roll attractor during inflation~\cite{Sugiyama_2013, Garriga:2016poh}. However this does not in general hold beyond slow roll, and we have shown explicitly by computing the isotropic part of the momentum constraint that it is not suppressed on large scales in ultra-slow-roll inflation. 

The separate-universe approach is a key ingredient in many non-linear analyses of cosmological perturbations, and in particular in the stochastic-inflation formalism, where fields are coarse-grained above some smoothing scale~\cite{Starobinsky:1986fx,Starobinsky:1994bd,Vennin:2015hra}. Quantum fluctuations within each patch then give a stochastic kick to the local evolution of the coarse-grained field as they are stretched beyond the smoothing scale. Quantum field fluctuations can be calculated using the Mukhanov-Sasaki variable \eqref{def:u} which describes the field fluctuations in the spatially-flat gauge and as we have seen this can lead to non-adiabatic field fluctuations in models of inflation beyond slow roll. As a consequence, the Hamilton-Jacobi approach implemented in Refs.~\cite{Prokopec:2019srf,Rigopoulos:2021nhv} fails to describe the full stochastic evolution in the spatially-flat gauge on large scales in ultra-slow roll and other models of inflation beyond slow roll. Instead one must allow for stochastic evolution in the full two-dimensional phase space, $H(\phi,\dot\phi)$, on super-Hubble scales~\cite{Firouzjahi:2018vet,Firouzjahi:2020jrj,Figueroa:2020jkf,Pattison:2021oen,Tomberg:2022mkt,Mishra:2023lhe,Jackson:2023obv,Ballesteros:2024pwn,Jackson:2024aoo}.

\acknowledgments
We would like to thank Antonio  Enea  Romano, Joseph Jackson, Atsushi Naruko, Misao Sasaki, Takahiro Tanaka, Eemeli Tomberg and Ashley Wilkins for interesting discussions. D.~A., E.~F.~and T.~M. are thankful to the University of Portsmouth for receiving them, where this work was initiated. D.~A., V.~V.~and D.~W.~are grateful to the Yukawa Institute for Theoretical Physics, Kyoto University, and the organisers of the Gravity and Cosmology 2024 workshop for their hospitality. D.~A.~is supported by JSPS Grant-in-Aid for Scientific Research No. JP23KF0247. This work was supported by the Science and Technology Facilities Council (grant number ST/W001225/1). For the purpose of open access, the authors have applied a Creative Commons Attribution (CC-BY) licence to any Author Accepted Manuscript version arising from this work. 

\paragraph{Note added.}
Ref.~\cite{Prokopec:2025uvz} appeared after this paper was accepted, where the authors propose a modification of the usual Hamilton-Jacobi approach during a period of ultra-slow roll.

\appendix
\addtocontents{toc}{\protect\setcounter{tocdepth}{1}} 
\section{Notations in the Lagrangian formalism}
\label{app:lag}
We here translate our notations in terms of the variables commonly used in the Lagrangian formalism. Namely, the perturbed metric \eqref{eq:metric:ADM} is written in conformal time as \cite{Malik:2008im}
\bea
	\dd s^2 &=& -a^2 \left(1+2 A\right) \dd \eta^2 + 2 a^2 \partial_{i} B \,\dd \eta \, \dd x^j + \, a^2 \left[\left(1 - 2\psi\right) \delta_{ij}+ 2 \partial_i \partial_j  E \right] \dd x^i \dd x^j \,.
\eea

A direct comparison with the notations used in \Eq{eq:metric:ADM} leads to
\bea
A\left(t,\vec{k}\right) &\equiv & \frac{\delta N \left(t,\vec{k}\right)}{N(t)}\,, \\
B\left(t,\vec{k}\right) &\equiv &  \frac{\delta N_1 \left(t,\vec{k}\right) }{k} \,,\\
\psi\left(t,\vec{k}\right) &\equiv & \frac{1}{a^2(t)} \left( - \frac{\delta \gamma_1 \left(t,\vec{k}\right)}{2\sqrt{3}} + \frac{\delta \gamma_2 \left(t,\vec{k}\right)}{2 \sqrt{6}} \right) \,,\\
E\left(t,\vec{k}\right) &\equiv & - \frac{1}{2} \sqrt{\frac{3}{2}} \frac{\delta\gamma_2 \left(t,\vec{k}\right)}{k^2 \, a^2(t)}\,.
\eea

For convenience, we also give the expression of the gauge-invariant Bardeen potential in both conventions
\bea
\Psi_{\mathrm{B}} &:= & \psi - a H \left(B - \dot{E}\right) \\
 &= & - \left( \frac{1}{2\sqrt{3}\, a^2} \right) \delta\gamma_1 + \left[ \frac{1}{2\sqrt{6} \, a^2} + \frac{1}{\Mp^2 k^2} \sqrt{\frac{2}{3}} \left( \frac{\pi_\phi^2}{2 a^6} +V \right) \right] \delta\gamma_2 \nonumber \\
&& + \left( \frac{1}{\Mp^4 k^2} \sqrt{\frac{3}{2}} a \theta \right) \delta\pi_2\,. \label{eq:BardeenConformal}
\eea

Let us show that the Bardeen potential is proportional to the non-adiabatic pressure perturbation. One can use the expression of $\delta\mathcal{D}$, \Eq{eq:mom}, in order to replace the $\delta \pi_1$ contribution in $\delta \mathcal{S}$, \Eq{eq:scal}. Such a combination of the two constraints boils down to
\bea
0 &=& \left(a^3 V_{,\phi} +3 \pi_\phi H \right) \delta\phi + \frac{\pi_\phi}{a^3} \delta\pi_\phi - \left(\frac{\sqrt{3}}{2} \frac{\pi_\phi^2}{a^5} + \frac{\Mp^2 k^2}{\sqrt{3} a} \right) \delta\gamma_1 \nonumber\\
&& + \left[ 2 \sqrt{\frac{2}{3}} \left( \frac{\pi_\phi^2}{2 a^5} + a V \right) + \frac{\Mp^2 k^2}{\sqrt{6} a} \right] \delta \gamma_2 -2 \sqrt{6} a^2 H \delta\pi_2 \,. \label{eq:HamJac}
\eea
Injecting \Eq{eq:HamJac} in the expression for the non-adiabatic pressure perturbation \eqref{eq:dPnad} in order to get rid of the $\delta\phi$ and $\delta\pi_\phi$ contributions, one can recast $\delta \Pnad$ into the form
\bea
\delta \Pnad &=& - \left(\frac{2 \Mp^2 k^2}{3\sqrt{3}} \frac{V_{,\phi}}{a \pi_\phi H} \right) \delta\gamma_1 + \left( 4 \sqrt{\frac{2}{3}} \Mp^2 \frac{a H V_{,\phi}}{\pi_\phi} + \frac{\Mp^2 k^2}{3}\sqrt{\frac{2}{3}} \frac{V_{,\phi}}{a\pi_\phi H} \right) \delta\gamma_2 \nonumber \\
&&- 4\sqrt{\frac{2}{3}} \frac{a^2 V_{,\phi}}{\pi_\phi} \delta\pi_2 \,. \label{eq:dPnadBis}
\eea
One can check explicitly that
\bea
\delta \Pnad =  \frac{4 \Mp^2 k^2}{3}  \frac{a V_{,\phi}}{\pi_\phi H} \Psi_{\mathrm{B}} \,,
\eea
where $\Psi_{\mathrm{B}}$ is the Bardeen potential in conformal time, as defined in \Eq{eq:BardeenConformal}.\footnote{This can also be recast in the form \bea
\delta \Pnad = - \frac{2 V_{,\phi}}{3H\dot{\phi}}  \frac{\nabla^2\Psi_{\mathrm{B}} }{4\pi G a^2} \,.
\eea}

One can also consider the usual extrinsic-curvature tensor. On a FLRW background, when restricting to scalar perturbations in the spatially-flat gauge, its traceless part reduces to
\bea
\delta A^i_{j} = \frac{1}{N} \partial_j \delta N^i \,, \qquad \forall \, i\neq j \,.
\eea
In Fourier space, this yields
\bea
\delta A^i_{j} = - \frac{k^i k_j}{a k} \delta N_1 = - \frac{k^i k_j}{k^2} \frac{\sqrt{6}}{a \Mp^2} \delta\pi_2 =  \frac{k^i k_j}{k^2} \frac{3 \pi_\phi}{4 a^3 V_{,\phi} \Mp^2} \delta\Pnad \,, \qquad \forall \, i\neq j \,,
\eea
in conformal time, where we used Eq.~(3.36) of \cite{Artigas:2021zdk} and Eq.~\eqref{eq:dPnadSF} for the second and third equalities respectively. Note that in the spatially-flat gauge, the non-adiabatic pressure perturbation is directly related to the traceless part of the extrinsic curvature.

\section{Slow-roll parameters} \label{App:SRParam}
The slow-roll parameters in conformal time are defined as
\bea
&\epsilon_1 &:= - \frac{H'}{a H^2} \,, \label{eq:epsilon1} \\
&\epsilon_{n+1} &:= \frac{1}{a H} \frac{\epsilon'_n}{\epsilon_n} \,.
\eea
A direct computation for the first and second slow-roll parameters leads to
\bea
\epsilon_1 &=&\frac{1}{2\Mp^2} \frac{\pi_\phi^2}{a^6 H^2} \,, \\
\epsilon_2 &=& -6 + 2\epsilon_1 - 2 \frac{a^3 V_{,\phi}}{H \pi_\phi} \,.
\eea
As a consequence,
\bea
\pi_\phi &=& \sigma \Mp a^3 H \sqrt{2\epsilon_1} \,, \\
V_{,\phi} &=& - \sigma H^2 \Mp \sqrt{\frac{\epsilon_1}{2}} \left(6+\epsilon_2 - 2\epsilon_1 \right) \,,
\eea
with $ \sigma:= \textup{sign}\left(\pi_{\phi}\right)$.

The scale factor can be expanded in terms of the slow-roll parameters (see \cite{Pattison:2019hef} for more detailed calculations). Below, $t$ refers to cosmic time. It relates to the conformal time as follows:
\bea
\eta := \int \frac{\dd t}{a} = - \frac{1}{aH} + \int \frac{\epsilon_1 \dd a}{a^2 H} \,.
\eea
After an integration by parts and an expansion in terms of the first slow-roll parameter, the integral boils down to
\bea
\eta = - \frac{1}{aH} - \frac{\epsilon_1}{aH} + \int \epsilon_1 \epsilon_2 \frac{\dd a}{a^2 H} + \mathcal{O}\left(\epsilon_1^2 \right) \,. \label{eq:eta}
\eea

\subsection{Slow-roll expansion} \label{App:SRExp}
In slow roll, both $\epsilon_1$ and $\epsilon_2$ are small. Therefore, at first order in slow-roll expansion the conformal time~\eqref{eq:eta} approximates to
\bea
\mathcal{\eta} = -\frac{1}{a H} \left(1 + \epsilon_1\right)\,.
\eea
The slow-roll parameters can be Taylor expanded and, when neglecting terms of order $\mathcal{O}\left(\epsilon_i^2\right)$, $\epsilon_1$ is approximately constant $\epsilon_1=\epsilon_{1*}$, where the $*$ refers to the Hubble-crossing time of the mode considered, $\eta_* = -1/k$. Inverting the above equation to integrate $H$, one eventually gets
\bea
\frac{a}{a_*} &=& \frac{\eta_*}{\eta} \left[1-\epsilon_{1*} \ln\left(\frac{\eta}{\eta_*}\right) \right] \,.
\eea
Taking \Eq{eq:eta} at time $\eta_*$ we get the initial condition $a_* \eta_* = - \left(1+\epsilon_{1*}\right)/H_*$. Plugging this identity in the expression for $a$ we get
\bea
a = - \frac{1}{H_* \eta} \left[1 + \epsilon_{1*} \left(1 - \ln\frac{\eta}{\eta_*}\right)\right]\,,
\eea
at first order in slow roll.

\subsection{Ultra-slow-roll expansion}  \label{App:USRExp}
In the USR case the potential is considered to be almost constant.
As a consequence $\epsilon_2 \approx -6 + 2\epsilon_1$ and terms of order $\epsilon_1\epsilon_2$ should not be neglected. Taking this into account, \Eq{eq:eta} can be approximated to
\bea
\eta &=& - \frac{1}{aH} - \frac{\epsilon_1}{aH} - 6 \int \frac{\epsilon_1 \dd a}{a^2 H} + \mathcal{O}\left(\epsilon_1^2 \right)\,,
\eea
where we, however, neglect terms of order $\mathcal{O}\left(\epsilon_1^2\right)$. The integral in the right-hand side can be integrated by parts, leading to
\bea
\int \frac{\epsilon_1 \dd a}{a^2 H} = - \frac{1}{7} \frac{\epsilon_1}{aH}\,.
\eea
The expression for $\eta$ therefore becomes
\bea
\eta &=& - \frac{1}{aH} \left(1 + \frac{\epsilon_1}{7} \right)\,.
\eea
In this paper, we will focus on the leading order in $\epsilon_1$ only and approximate $\eta=-1/(aH)$.

\bibliographystyle{JHEP}
\bibliography{dPnad}

\end{document}